\documentstyle[12pt]{article}

\begin{document}

\pagestyle{plain}

\parskip0.3cm
\parindent0.75cm

\begin{flushright}
\today
\end{flushright}

\begin{center}
{\Large \bf Bundles of interacting strings in two dimensions}\\[3ex]
{\bf Christin Hiergeist, Michael L\"assig and Reinhard Lipowsky} \\[1ex]
{\sl Institut f\"ur Festk\"orperforschung \\
Forschungszentrum J\"ulich, 52425 J\"ulich, Germany\\
and\\
Max--Planck-Institut f\"ur Kolloid-- und Grenzfl\"achenforschung,\\
Kantstr. 55, 14513 Teltow--Seehof, Germany}
\end{center}

\begin{abstract}

Bundles of strings which interact via short-ranged pair potentials are studied 
in two dimensions. The corresponding transfer matrix problem is solved 
analytically for arbitrary 
string number $N$ by Bethe ansatz methods.
Bundles consisting of $N$ identical strings exhibit a unique unbinding 
transition. 
If the string bundle interacts with a hard wall, the bundle may unbind from the 
wall via a unique transition or a sequence of $N$ successive transitions. 
In all cases, the critical exponents are independent of $N$ and the density
profile of the strings exhibits a scaling form that approaches 
a mean-field profile in the limit of large $N$.
\\
\\
\noindent PACS numbers: 68.10, 64.70, 82.70-y

\end{abstract}

In the context of condensed matter physics, strings are essentially
1--di\-men\-sional objects which are (i) directed in the sense that their 
tangent
vectors point, on average, into a certain direction, and (ii) are governed by a
finite line tension. Physical examples are domain walls in adsorbed
monolayers \cite{pokr80}, steps or ledges on crystal surfaces \cite{grub67},
vortex lines in type--II superconductors \cite{fett69}, stretched polymers
\cite{genn68} and presumably some polyelectrolytes \cite{kan89}. 
Two different ensembles of strings have to be distinguished: (i) systems with a
fixed density of strings and (ii) systems with a fixed number $N$ of strings,
which are the topic of this letter. If the strings have attractive
interactions, they may at low temperatures be bound together to a bundle.
Such bundles of strings have been studied 
by numerical diagonalization of the
transfer matrix \cite{netz93a,netz94}, 
in a local density functional theory \cite{helf93}, 
by mapping onto a quantum spin chain 
\cite{burk93},
in a heuristic scaling picture
\cite{lip94}, and by field-theoretic renormalization group methods \cite{laes}.

In this paper, we study bundles of 
$N$ strings which interact via
contact
pair potentials. 
Such a system of strings can be mapped onto a system of $N$
quantum--mechanical particles interacting via the same pair potentials. 
For $N$ strings in two dimensions, one is then led to consider a
Schr\"odinger--type equation in $N$ dimensions 
which can be solved analytically for {\it arbitrary} $N$ 
using the Bethe product ansatz (see e.g. \cite{thac81} for a review).
Often, real strings do not intersect which 
can be taken into account by imposing Fermi statistics on the particles 
\cite{genn68,pokr78}. Here, we take a different avenue. We modify the 
contact interaction so as to
impose a {\em preferred ordering} on the strings while {\em preserving
integrability}. Hence we construct a {\em one-parameter family} of Bethe ansatz 
solutions that interpolate between intersecting and non-intersecting strings.

In this way, we consider two cases: 
(i) Free bundles consisting of $N$ identical strings
interacting via identical pair potentials. For this case, we find
that the bundle undergoes a unique unbinding transition which is
characterized by universal, i.e., $N$--independent critical exponents.  We also
calculate the density profile of the strings for $N$ = 2, 3 and 4 and within a
mean--field approximation. For large $N$, the density profile seems to
converge towards the mean--field profile, see Fig.\ 1(a) below.
(ii) Bundles interacting with a rigid wall. Extending the 
corresponding Bethe ansatz for 
intersecting strings due to Kardar \cite{kar87},
we find a complex phase diagram, see Fig.\ 2 below. Depending on the
relative strength of the string--string and the string--wall interaction, the
bundle may unbind from the wall via a unique transition or via a sequence of 
transitions. In
all cases, the critical exponents are universal, i.e., independent of $N$.

The free bundle consists of 
$N$ strings with identical 
stiffness $K$. The strings are infinitely extended and run, on average, 
parallel to the 
$x$--direction. Their configurations  
are parametrized by 
the displacement fields $l_n(x)$ with $n=1,\ldots,N$.
The effective Hamiltonian for the bundle is 
given by
\begin{equation} 
{\cal H}\{l_1(x), \ldots ,l_N(x)\}= \int\!\!dx \{\frac{1}{2} K 
\sum_{n=1}^{N} (\frac{dl_n}{dx})^{2} + \sum_{i>j} V(l_i-l_j) \} 
 \hspace{2mm}
\end{equation}
where  $V(l_i-l_j)$ is the interaction potential for the string 
pair with $n=i$ and $n=j$.
Since this model is 1--di\-men\-sional it can be studied by transfer matrix 
\mbox{methods}. 
In the limit of vanishing small distance cutoff, one obtains a  
Schr\"o\-ding\-er--type equation 
$
\hat{H} \psi(\{l_n\}) =
E \psi(\{l_n\})  
$
with the Hamilton operator
\begin{equation}
\hat{H} = 
-\frac{T^2}{2 K} \sum_{n=1}^{N} \frac{\partial^{2}}{\partial l_{n}^{2}}
  + \sum_{i>j} V(l_{i}-l_{j})
\hspace{2mm}. 
\label{nham}
\end{equation}

The strings interact with an attractive contact 
interaction 
\begin{equation}
V_0(l) \equiv - v_0   \delta (l) \hspace{2mm} {\rm with} 
\hspace{2mm}  v_0 > 0 
 \hspace{2mm}.
\end{equation}
In order to model the behaviour of non-intersecting strings the potential 
\begin{equation}
V_1(l) \equiv
       \frac{1}{\epsilon} v_1 \delta (l+ \epsilon) 
      -\frac{1}{\epsilon}\frac{v_1}{1+v_1 K/T^2} \delta (l)  
\quad {\rm with}
\quad v_1 > 0  
\end{equation}
is added. This additional potential does not
supress all intersections, but favours a specific ordering 
of the strings.
Therefore $V_1$ leads to 
partially intersecting strings. The parameter $v_1$ controlls the degree
of asymmetry in the probability distribution of the separation variables 
$l_i - l_j$.
In the limit of zero $\epsilon$, 
the potential $V_0 +V_1$ leads to a pair of 
matching conditions for the 
wavefunction $\psi$ and its first derivative $\psi'$. Both
functions are discontinuous at $l_i - l_j=0$; the height of the jump 
depends on
$g_0 \equiv v_0 K /T^2$ and $g_1 \equiv v_1 K /T^2$.

Solving the Schr\"o\-ding\-er equation for these 
matching conditions yields a localized ground state. 
The 
$N$--particle wavefunction is obtained 
by the simple product ansatz
\begin{equation}
\psi_0(l_1, \ldots ,l_N) \sim \prod_{i>j}(1+g_1 \theta(l_i -l_j)) 
                                       \exp(-p | l_i -l_j |)
\label{nwf1} \hspace{2mm}
\end{equation}
with the transverse momentum
\begin{equation}
p \equiv g_0 \frac{(1+ g_1)^2}{1+(1+ g_1)^2} 
\label{rho} 
\hspace{2mm}.
\end{equation}
For $g_1=0$, one recovers the groundstate for $N$ intersecting 
strings \cite{thac81}. The preferred ordering 
imposed to the strings for $g_1>0$ is
$l_1< \cdots < l_N$. In the limit of infinite $g_1$, the wavefunction 
vanishes for all other string permutations resembling, thus, the 
wavefunction of non--intersecting strings.

The free energy per unit length, 
$f(N)=E_0$, is given by
\begin{equation}
f(N) = - \frac{1}{6} N (N^2 -1) p^2 \frac{T^2}{K}
\approx - \frac{1}{6} N (N^2 -1) v_0^2
  \label{energ}  \hspace{2mm},
\end{equation}
where 
the asymptotic equality holds in the limit of large $v_1$, i.e., of 
non--intersecting strings as 
follows from (\ref{rho}). 
This expression for the free energy agrees with the 
result in \cite{burk93}.

We now introduce a new set of variables $\{\ell_1, \ldots ,\ell_N\}$.
For a given permutation $\sigma$ of the strings with
$l_{\sigma(1)}< \cdots < l_{\sigma(N)}$, $\ell_n $ is given by
$\ell_n \equiv l_{\sigma(n)}$.
In the limit of large~$v_1$, i.e., of 
non--intersecting strings, the mean position
$\langle \ell_n \rangle$ is equal to the mean position 
in the original variables,  $\langle l_n \rangle$.
The wavefunction~(\ref{nwf1}) is translationary
invariant;
the mean position $\langle \ell_n \rangle$ is therefore calculated
keeping $ \ell_1 =0$ fixed.
The mean extension of the bundle 
is 
then given by 
\begin{equation}
\ell_{bu} \equiv \langle\ell_N\rangle - \langle\ell_1\rangle 
\approx \ln (N) / N p  \quad \mbox{for large $N$}
\quad.  \label{mad}
\end{equation}
The mean separation between neighbouring strings behaves as
\begin{equation}
\Delta \ell_n \equiv \langle \ell_{n+1}\rangle - \langle\ell_n\rangle
= (2 p (N-n) n)^{-1} 
\label{mab2} \hspace{2mm},
\end{equation}
with $n=1,\ldots,N-1$.  
Both mean separations (\ref{mad}) and (\ref{mab2})
are characterized by the critical behaviour
\begin{equation}
\ell_{bu} \sim \Delta \ell_n \sim p^{-\psi}  \quad \mbox{with $\psi=1$}
\quad. 
\label{pdef}
\end{equation}
The continuum description 
used here is justified as long as the mean separation 
between the inner strings
is greater than the string thickness $a_{\perp}$ 
which implies $N \ll \sqrt{2/a_{\perp} p}$.

The string density
$
\rho_N(l) \equiv 
\langle \psi_0|\sum_{n=1}^N \delta(l-l_n) | \psi_0 \rangle 
$
has been calculated for
$N=2,3,4$ (where 
the center of mass coordinate was set equal to zero), see Fig.~1(a).
The results can be written in the scaling form
$
\rho_N(l)=2 p N^2 \, \Omega_N(2p N l) 
$
with the scaling function
\begin{eqnarray}
\Omega_N(z) =  \sum_{j=1}^{N-1} a_{N,j} \exp(-j|z|)
\end{eqnarray}
where $a_{N,1}=(N-1)/N$, $a_{3,2}=-1/3$,
$a_{4,2}=-3/5$, and $a_{4,3}=3/20$.
Note that $\int \! dl \, \rho_N(l) =N$ implies 
$\int \! dz \, \Omega_N(z) =1$.
The mean--field density
defined by
$
\rho_{MF}(l= \langle\ell_n\rangle - {\textstyle \frac{1}{2}} \ell_{bu}) 
\simeq 1/\Delta \ell_n  
$
has, in the limit of large $N$, an analogous scaling form with the
scaling function
\begin{equation}
\Omega_{MF}(z)
 =\left(2 \cosh (z/2)\right)^{-2}
\label{omegamf}  
\hspace{2mm}.
\end{equation}
This mean--field density is identical with the density 
obtained by
Helfrich in \cite{helf93}.
For large $N$, the exact densities $\Omega_N(z)$
seem to 
converge towards the  mean--field profile, see Fig.~1(a).

The unbinding of $N$ strings can be understood heuristically in the framework
of a $N$--state model \cite{lip94}. If one makes the
plausible assumption that locally bound triplets 
and higher--order multiplets of strings are less likely than 
locally bound pairs, one finds that the unbinding of the bundle is governed 
by the unbinding of string pairs. This explains that the
unbinding temperature does not depend on $N$ as has been observed in numerical 
studies for $N=3$ and $N=4$ \cite{netz94,netz93a}.
On the other hand, these numerical studies 
lead to an effective critical exponent $\psi$ which is nonuniversal and 
depends on $N$.
The total interaction potential studied numerically contains only 
contributions from 
neighbouring pairs of strings. In contrast the total interaction potential 
$\sum_{i>j} V(l_i-l_j)$ studied here contains contributions from {\it all}
pairs. The difference corresponds to a 3--string interaction which must be
added to $\sum_{i>j} V(l_i-l_j)$ and which is effectively repulsive. 
Such an interaction represents a marginally irrelevant perturbation 
and therefore leads to large corrections to scaling which should explain the
$N$-dependence found in the numerical work \cite{laes}.

Next, consider
the unbinding of the string bundle from a wall.
The $N$--string system 
now experiences an additional external 
potential 
consisting of an attractive well and a hard wall. This potential
causes the wavefunction of the adjacent string to fall off as
$\exp (-q \ell_1)$ (the possibility that more than one string is in the well 
is ignored).
The ground state wave function which is analogous to the
solution for intersecting strings \cite{kar87} is
of the
form
\begin{eqnarray}
\psi_0 \sim \prod_{i>j}(1+g_1 \theta(l_i -l_j)) 
                              \prod_n \exp(-p_n^{(w)} \ell_{n}) 
                       \label{nwfct} 
\end{eqnarray}
with the transverse momenta
$
p_n^{(w)} \equiv q + 2 (n-1) p \label{rhow}
$
where $p$ is still given by
(\ref{rho}). 

The free energy per unit length of $N$ strings bound to the wall is
\begin{equation}
f^{(w)}(N) = E_0^{(w)} =-\frac{1}{2} N (q + (N-1) p)^2 \frac{T^2}{K} + 
f(N)
\label{energw}   \hspace{2mm},  
\end{equation}
where $f(N)$ is the free energy per unit length
of the free bundle (\ref{energ}). 
The analysis 
shows that $f^{(w)}(N)=f(N)$ for $q=-(N-1)p$ and
$f^{(w)}(N)=f^{(w)}(N-1)$ 
for $q=-2(N-1)p$.
The mean separation between neighbouring strings is 
$
\Delta \ell_n
= [2 (N-n) (q+(N-1+n)p)]^{-1} 
$.

The state of the system is determined by three parameters: 
(i) the parameter $p$ of the string--string interaction; 
(ii) the transverse momentum $q$ resulting from the string--wall interaction; 
and (iii) the total number $N$ of 
strings. In Fig.~2 
the phase diagram is displayed for $N=5$. 
In regime ($\rm{B_N}$) given by (i)
$q> -(N-1) p$ for $ p>0$ and (ii) $q> -2(N-1) p$ for $p<0$, 
all $N$ strings are bound to the wall. 
For $p>0$, all $N$ strings unbind simultaneously 
from the wall at $q = -(N-1) p$. In regime (FB) with 
$q< -(N-1)p$ and $p>0$
the strings form a free (or unbound) 
bundle. 
For $p<0$, 
the strings 
unbind successively. 
The $n$th string (counted from the wall) peels from the wall at $q= -2(n-1)p$.
For $p<0$ and $0<q<-2(N-1) p$, we therefore find $N-1$ different 
regimes $\rm{(B_n)}$
with $1 \le n \le (N-1)$ strings bound to the wall and the remaining strings 
completely
unbound. 
For 
$q< 0$ and $p<0$ one has the free string (FS) regime.
When the point $p=q=0$ is approached 
from regime ($\rm{B_N}$) the strings unbind simultaneously from the wall 
as well as from each other.

Across all phase boundaries, both the free energy
$f^{(w)}$ and its
first derivative $\partial f^{(w)} / \partial q $ are continuous 
whereas the second derivative $\partial^2 f^{(w)} / \partial q^2 $ 
exhibits a discontinuity. Hence  these transitions are of second order
with the critical exponent $\alpha=0$ for the specific heat. 
At the phase boundary between ($\rm{B_N}$) and ($\rm B_{N-1}$), the
mean separations $\Delta \ell_n$ with $n<N$ are continuous but their
first derivatives $\partial \Delta \ell_n / \partial q $ exhibit a
jump.
The critical behaviour of the diverging length scales is the same
at each transition line, even at the point $p=q=0$, with the $N$-independent 
exponent $\psi=1$.

In a real system, regime ($\rm{B_N}$) and regime (FS) are attained at 
sufficiently low and sufficiently high temperatures, respectivly. 
Depending on the relative strength of the string--string and the 
string--wall attraction,
the temperature trajectory will move from ($\rm{B_N}$) to (FS) via
the free bundle regime (FB) or the intermediate 
states ($\rm{B_n}$). In the latter case one has
a sequence of $N$ unbinding transitions. In the
limit of infinite $N$, the sequence of critical tempertures $T_c(n)$ 
attains a finite value $T_c(\infty)$ \cite{netz93a}. It now follows from
the explicit expressions for the phase boundaries that 
\begin{equation}
1-(T_c(\infty)/T_c(n))^2 \sim 1/n^{\lambda} \quad \mbox{with} \quad 
\lambda=1
\hspace{2mm}.
\end{equation}

For a bundle in regime ($\rm{B_N}$) the exact 
density has the scaling form
$
\rho_N^{(w)}(l)=2p N^2 \, \Omega_N^{(w)} (2 p N  l, q/p(N-1)) 
$.
The scaling functions $\Omega_N^{(w)} (z,y)$
are displayed in Fig.~1(b) for $2 \le N \le 7$ and $y=-0.95$. Note, that
the phase boundary between regime ($\rm{B_N}$) and regime (FB) is
located at $q=-(N-1)p$ and, thus, at $y_c=-1$.  
The mean--field density 
which is defined by 
$\rho_{MF}^{(w)}(l= \langle\ell_n\rangle) \simeq 1/\Delta \ell_n $ 
has the scaling form 
$
\rho_{MF}^{(w)}(l)=2 p N^2 \, \Omega_{MF}^{(w)} (2 p N l, q/N p) 
$
with
\begin{equation}
\Omega_{MF}^{(w)}(z,y)=(1+y/2)^2 
           \left[ \cosh \left( (1 + y/2) (z - z_{max}) \right)
         \right]^{-2}
     \label{mdw}  
\end{equation}
and $z_{max} \equiv -(2+y)^{-1} \ln (1+y) $.
The unbinding transition towards the (FB) regime is not correctly 
described by the mean--field density. As $y$ approaches $y_c=-1$ from above, 
the mean separation of the bundle from the wall scales as 
$ \sim \ln (1/|y-y_c|)$ within the mean--field theory whereas the 
correct critical 
behaviour is given by $\sim |y-y_c|^{-\psi}$ with $\psi=1$.

The transition from regime ($\rm{B_N}$) to regime ($\rm B_{N-1}$) occurs at
$p=-q/2(N-1)$. In this case, the mean field density 
exhibits a power law tail which is given by
\begin{equation}
\rho_{MF}^{(w)}(l) \approx N l_{sc}/(l_{sc} + l)^2 
\quad \mbox{for large $N$}
\label{mfdenpt} 
\end{equation}
where $l_{sc} \equiv 1/q$.
As before the continuum limit is only justified as long as the string
separation $\Delta \ell_n$ is larger than the string thickness $a_{\perp}$. 
This leads to the 
crossover scale $l_{\ast}$ defined by $\rho_{MF}^{(w)}(l=l_{\ast})=1/a_{\perp}$.
The strings with $l<l_{\ast}$ are densely packed; 
the 
strings with $l>l_{\ast}$ are swollen and should exhibit the power law tail 
as 
in (\ref{mfdenpt}).
The swollen 
region contains $n_{sw} \sim (N/q a_{\perp})^{1/2}$ 
strings.

In summary, we have obtained analytic results on the unbinding transition 
and density profiles of bundles of
$(1+1)$-dimensional strings 
both for a free bundle and for a bundle interacting with a rigid wall.
The critical behaviour we have found here should apply to all 
pair potentials which decay faster than $  l^{-2}$ for large $l$ and hence
belong to the strong--fluctuation regime. 
In
general, many--string forces are present as well; for example, 
the force between
two strings may be screened by a third string in between. 
Renormalization group
arguments \cite{laes} as well as the scaling picture of \cite{lip94} show that
such screening forces do not alter the asymptotic scaling.
With
large {\em attractive} many-string forces, however, the transition is governed
by a different fixed point, and the Bethe  ansatz breaks down \cite{laes}.
Further, the 
critical behaviour of fluid membrane bunches should be analogous to the 
behaviour 
of string bundles. Therefore, the asymptotic critical behaviour for $N$ 
identical membranes 
should be governed by universal critical exponents. 
Another interesting problem are strings interacting 
via short--ranged pair potentials in $d=1 + d_{\perp}$ dimensions. 
For $d_{\perp} <4$, two strings exhibit continuous unbinding transitions 
\cite{lip91}, and one would expect that the associated critical behaviour also 
applies to string bundles with $N>2$. An explicit calculation for this system 
would be quite valuable.

\section*{Acknowledgements}
 
We thank Frank J\"{u}licher and Roland Netz for stimulating interactions.

\section*{Figure Captions}

\begin{description}
\item[Fig.~1(a)] 
Density profiles for a free bundle of strings: The 
exact densities $\Omega_N \sim \rho_N$ 
as a function of separation $z \sim l$ for $N=2,3,4$ strings 
together with the mean--field density $\Omega_{MF}$.
For increasing $N$, the exact densities seem to converge 
towards the mean--field profile. 
 
\item[Fig.~1(b)] 
Density profile for a string bundle interacting with a wall: The 
exact densities $\Omega_N^{(w)}$ as a function of separtion 
$z \sim l$ for
$2 \le N \le 7$ together with the mean--field profile 
$\Omega_{MF}^{(w)}$ for $y=-0.95$. The transition from the adhering 
bundle, regime ($\rm{B_N}$), to the free bundle regime (FB) is
located at $y_c=-1$. 

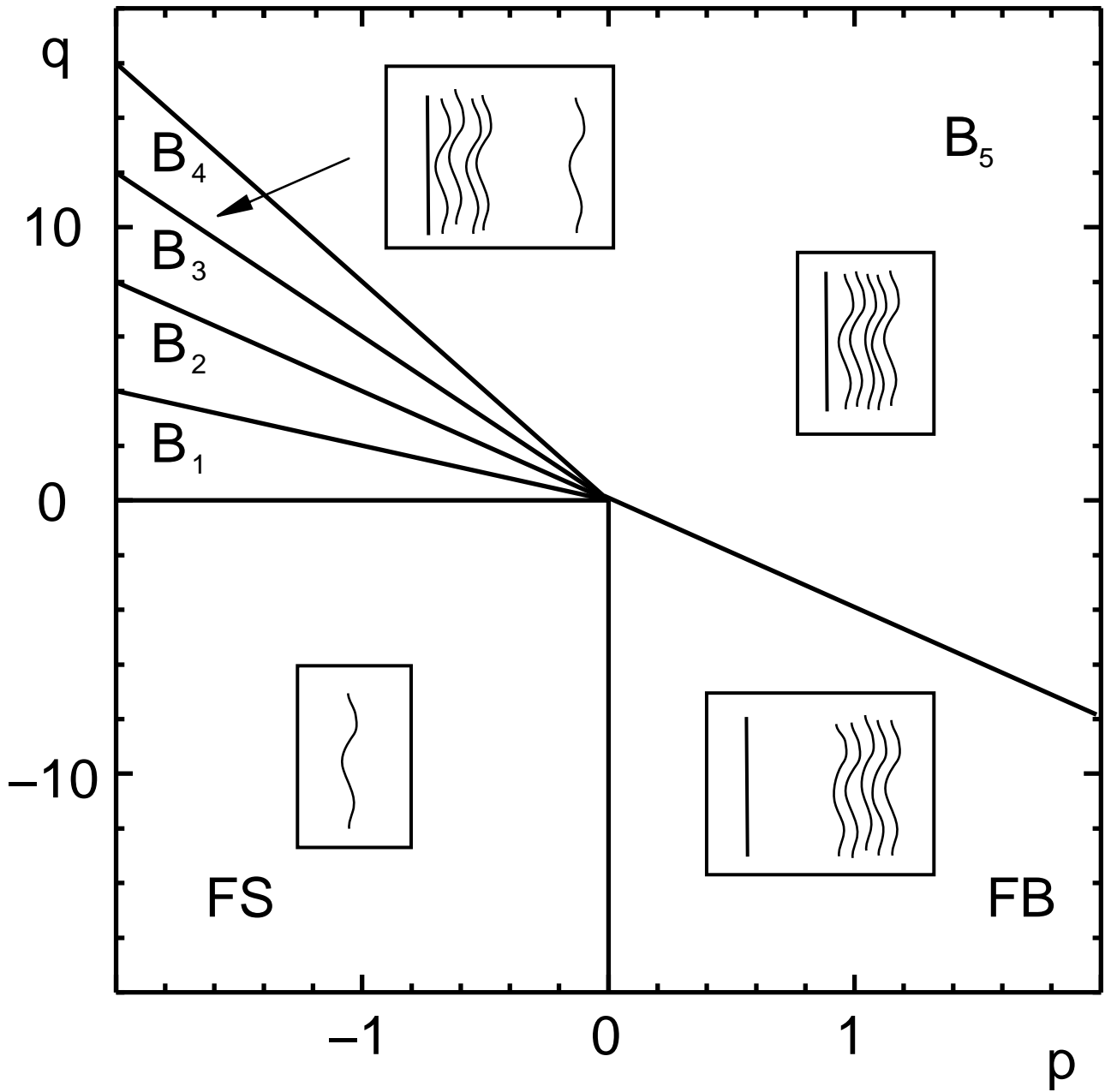
\item[Fig.~2] 
Phase diagram for a bundle of $N=5$ strings interacting with a wall. 
The parameter $p$ measures the string--string interaction, 
the parameter $q$ the string--wall interaction. 
In regime ($\rm{B_n}$), $n$ strings are bound to the wall with $N-n$ 
strings diffusing freely. 
In the free bundle regime (FB) the bundle is unbound with respect to the wall.  
The strings are completely unbound in the free string regime (FS).

\end{description}

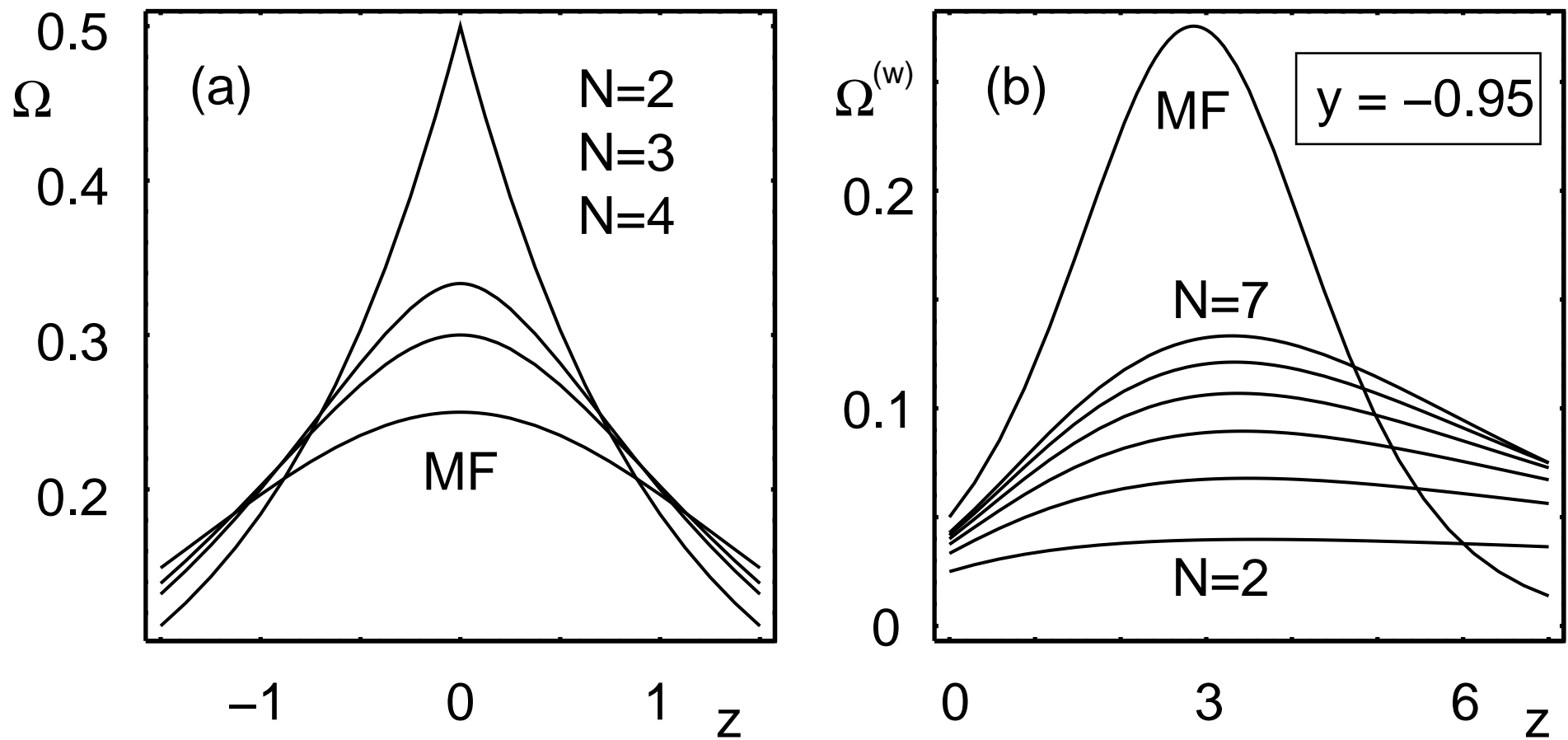
\end{document}